\documentclass{article}

\usepackage{epsfig}

\begin{document}

\title{The generalized quantum isotonic oscillator}

\author{J Sesma\footnote{javier@unizar.es}\\  \   \\
Departamento de F\'{\i}sica Te\'{o}rica, \\ Facultad de Ciencias,
\\ 50009 Zaragoza, Spain. \\  \ }

\maketitle

\begin{abstract}
Recently, it has been proved that a nonlinear quantum oscillator, generalization of the isotonic one, is exactly solvable for certain values of its parameters. Here we show that the Schr\"odinger equation for such an oscillator can be transformed into a confluent Heun equation. We give a very simple and efficient algorithm to solve it numerically, no matter what the values of the parameters are. Algebraic quasi-polynomial solutions, for particular values of the parameters, are found.
\end{abstract}


\section{Introduction}

Two years ago, Cari\~nena, Perelomov, Ra\~nada and Santander \cite{cari} considered a quantum oscillator, intermediate between the harmonic and the isotonic ones, whose Schr\"odinger equation reads
\begin{equation}
\frac{d^2}{dx^2}\Psi - \left[\omega^2x^2+2g\frac{x^2-a^2}{(x^2+a^2)^2}\right]\Psi+2E\,\Psi=0.  \label{uno}
\end{equation}
The interest of those authors on that problem lay on the fact that, as they proved, it is exactly solvable for certain values of the parameters, namely, $g=2$, and  $\omega$  and $a$ such that $\omega a^2=1/2$. Very recently, Fellows and Smith \cite{fell} have shown that this particular case of generalized isotonic oscillator is a supersymmetric partner of the harmonic oscillator. This fact has allowed them to reproduce the results of Ref. \cite{cari} in a very concise and elegant manner, and even to construct an infinite set of oscillators, with potentials approaching that of an isotonic oscillator as $x\to\infty$, all of them being partners of the harmonic oscillator and, therefore, exactly solvable. Related also to the generalized isotonic oscillator of Cari\~nena {\it et al.}, another class of exactly solvable problems has been obtained by Kraenkel and Senthilvelan \cite{krae}. They have used point canonical transformations to convert Eq. (\ref{uno}) into a Schr\"odinger equation with a position dependent effective mass, which, with adequate mass distributions, may represent different problems encountered in semiconductor physics. On the other hand, in view of the considerable progress in the synthesis of artificial quantized structures, a great variety of shapes for the potential wells and barriers are easily feasible and it seem plausible to see generalized isotonic oscillators as possible representations of realistic quantum dots.

The rising interest on generalizad isotonic oscillators has lead us to try to contribute to a better understanding of their main features. Dealing with Eq. (\ref{uno}) without any restriction, apart from the trivial ones
\[
\omega > 0, \qquad g> 0, \qquad a^2> 0,
\]
 about the values of the parameters,
we have found that a very natural change of variable transforms it into a confluent Heun equation (CHE). Then, the energies $E$ can be easily obtained as the zeros of a function defined by a continued fraction. Of course, the results of Ref. \cite{cari} are reproduced by our procedure. This allows us to relate, by a continuous variation of the parameter $g$,  the energy levels of the harmonic oscillator ($g=0$) with those of the generalized isotonic oscillator. Besides this, quasi-polynomial (i. e. product of a rational function times an exponential times a polynomial) wavefunctions appear for specific energy levels and particular values of the parameters $\omega$, $g$ and $a$, different from those considered in Ref. \cite{cari}.

In Section 2 we transform the Schr\"odinger equation (\ref{uno}) into a CHE.  To solve it, we propose an extremely simple algorithm that allows us to obtain the eigenvalues and the eigenfunctions with a great accuracy. Results of that algorithm are shown in Section 3. quasi-polynomial solutions of the Schr\"odinger equation are obtained in Section 4.  The relation of the polynomials found by Cari\~nena {\it et al.} \cite{cari} with confluent Heun polynomials is discussed in Section 5. Some final comments are added in Section 6.

\section{A confluent Heun equation}

Equation (\ref{uno}) presents two regular singular points, at $x=\pm ia$, and an irregular one of $s$-rank 3 at infinity. (Along this paper we adopt the definition of $s$-rank of an irregular singular point given in Sec. 1.2 of the book by Slavyanov and Lay \cite{slav}.) Taking advantage of the symmetry of the potential, the number and/or rank of the singularities can be reduced by a very natural mapping, namely,
\begin{equation}
z=x^2/a^2\,. \label{dos}
\end{equation}
In this way, the upper half of the $x$-plane goes into the whole $z$-plane and the real axis $x\in (-\infty, +\infty)$, relevant from the physical point of view, is mapped into the positive real semiaxis in the $z$-plane, covered  from $\infty$ to $0$ along the ray $\arg z=2\pi$ and from $0$ to $\infty$ along the ray $\arg z=0$. The Schr\"odinger equation turns into
\begin{equation}
\frac{d^2\Psi}{dz^2}  + \frac{1}{2z}\,\frac{d\Psi}{dz} + \left[-\frac{\omega^2a^4}{4}-\frac{g}{2}\,\frac{z-1}{z(z+1)^2}+\frac{Ea^2}{2z}\right]\Psi=0,  \label{tres}
\end{equation}
where the two free parameters $\omega a^2$ and $g$ are assumed to be given, whereas $Ea^2$ represents the eigenvalues to be determined.
That equation has two regular singularities, at $-1$ and $0$, and an irregular one of $s$-rank 2 at infinity. This suggests to compare it with some one of the various forms of the CHE, which presents the same pattern of singularities. Extensive discussions of the CHE can be found in Refs. \cite{slav} and \cite{ronv}. Equation (\ref{tres}), when written in the form
\begin{equation}
\frac{d^2\Psi}{dz^2}  + \frac{1}{2z}\,\frac{d\Psi}{dz} + \left[-\frac{\omega^2a^4}{4}+\frac{g/2}{z}-\frac{g/2}{z+1}-\frac{g}{(z+1)^2}+\frac{Ea^2}{2z}\right]\Psi=0\,,  \label{cuatro}
\end{equation}
is an example of CHE in its natural form \cite[Eq. (1.1.4)]{ronv}. The singularity at $z=-1$, coming from the singularity at $x=ia$, has indices
\begin{equation}
 \rho_1=\frac{1}{2}\left(1+\sqrt{1+4g}\right), \qquad \rho_2=\frac{1}{2}\left(1-\sqrt{1+4g}\right).   \label{cinco}
 \end{equation}
 The mapping (\ref{dos}) has introduced the singularity (branch point) at $z=0$, with indices
 \begin{equation}
 \nu_1=0,\qquad \nu_2=1/2\,,  \label{seis}
 \end{equation}
 which correspond, respectively, to even and odd solutions $\Psi (x)$. The singularity for $z\to\infty$ corresponds to that for $x\to\infty$. It is immediate to check that the two {\it Thom\'{e}} formal solutions of (\ref{tres}) for $z\to\infty$ behave as
 \begin{equation}
\exp\left(\pm\,\frac{\omega a^2}{2}\,z\right)z^{\mu_\pm}\,\left(1+O(z^{-1})\right), \qquad \mbox{with}\quad \mu_\pm=\mp\frac{E}{2\omega}-\frac{1}{4}\,.  \label{siete}
\end{equation}
The problem of finding the energy levels of the generalized isotonic oscillator reduces, thus, to determine the values $Ea^2$ such that the {\it Floquet} solutions (series of increasing powers of $z$) of (\ref{tres}), corresponding to the indices $\nu_1$ or $\nu_2$, vanish at infinity according to (\ref{siete}) with the minus sign in the argument of the exponential. This is the well known {\it connection problem} of the singular points at $0$ and at $\infty$. The presence of the singular point at $z=-1$ prevents the convergence of the Floquet solutions out of the unit disc. So, a process of analytic continuation would be necessary to find their behaviour for $z\to\infty$. Instead of this, Slavyanov and Lay \cite[Sec. 3.6]{slav} suggest to carry out a {\it Jaff\'{e} transformation} consisting in an adequate linear transformation of the dependent variable  followed by a M\"obius transformation of the independent variable to convert the interval $z\in [0,\infty)$ into the interval $[0,1]$ for the new variable. We have found convenient to substitute
\begin{equation}
\Psi(z) = (z+1)^\mu\,\exp\left(-\,\frac{\omega a^2}{2}\,z\right)\,w(z)\,,\qquad \mbox{with}\quad \mu=\mu_-=\frac{E}{2\omega}-\frac{1}{4}\,,   \label{ocho}
\end{equation}
in (\ref{tres}) to get
\begin{eqnarray}
 \lefteqn{\frac{d^2w}{dz^2}  + \left(-\omega a^2+\frac{1}{2z}+\frac{2\mu}{z\! +\! 1}\right)\,\frac{dw}{dz}} \nonumber  \\ 
  & & \mbox{} + \left[\frac{\mu(\omega a^2\! +\! 1/2)}{z(z+1)}+\frac{\mu(\mu\! -\! 1)}{(z+1)^2}-\frac{g}{2}\,\frac{z-1}{z(z\! +\! 1)^2}\right]w=0,  \label{nueve}
\end{eqnarray}
and then to apply the M\"obius transformation prescribed by Slavyanov and Lay
\begin{equation}
t=\frac{z}{z+1}\,,   \label{diez}
\end{equation}
which transforms Eq. (\ref{nueve}) into (keeping the same symbol to represent the dependent variable in terms of the new independent one)
\begin{eqnarray}
\lefteqn{\frac{d^2w}{dt^2} + \left(-\frac{\omega a^2}{(1-t)^2}+\frac{1}{2t(1-t)}+\frac{2\mu-2}{1-t}\right)\,\frac{dw}{dt}} \nonumber  \\
 & & \mbox{} + \left(\frac{\mu(\omega a^2+1/2)}{t(1-t)^2}+\frac{\mu(\mu-1)}{(1-t)^2}-\frac{g}{2}\,\frac{2t-1}{t(1-t)^2}\right)w=0\,,  \label{duno}
\end{eqnarray}
to be solved in the interval $t\in[0,1]$. The energy levels are the values of $E$ such that the ``even" ($\nu=\nu_1=0$) or ``odd" ($\nu=\nu_2=1/2$) series solution
\begin{equation}
w(t) = t^\nu\,\sum_{n=0}^\infty c_n\,t^n \,,  \qquad c_0\neq 0  \label{ddos}
\end{equation}
becomes finite at $t=1$. Substitution of (\ref{ddos}) in (\ref {duno}) gives for the coefficients $c_n$ the recurrence relation
\begin{equation}
A_{n+1}\,c_{n+1} + B_n\, c_n + C_{n-1}\,c_{n-1} = 0,   \label{dtres}
\end{equation}
where we have abbreviated
\begin{eqnarray*}
A_m &=& (m+\nu)(m+\nu-1/2)\,,  \\
B_m &=& (m+\nu)(-2m-2\nu-\omega a^2 +2\mu-1/2) +\mu(\omega a^2+1/2) +g/2\,,  \\
C_m &=& (m+\nu)(m+1+\nu-2\mu)+\mu(\mu-1)-g\,.  
\end{eqnarray*}
The recurrence relation (\ref{dtres}) is an irregular difference equation of the Poincar\'{e}-Perron type. The procedure suggested by Slavyanov and Lay to solve the connection problem consists in considering the physically acceptable solution of (\ref{dtres})  as  a linear combination of its two Birkhoff solutions and adjust the value of $Ea^2$ so as to reach the cancelation of the coefficient of the exponentially divergent one. Instead of this, we have preferred to base ourselves on the algorithms proposed by Gautschi \cite{gaut} to find minimal solutions of three-term recurrence relations. The crucial consideration is that the solution minimal for $n\to\infty$ turns out to be dominant when the recurrence is used ``from tail to head". Bearing this in mind, we write the above recurrence relation in the form
\begin{equation}
\frac{c_{n-1}}{c_n}=-\frac{B_n}{C_{n-1}}-\frac{\displaystyle{\frac{A_{n+1}}{C_{n-1}}}}{\displaystyle{\frac{c_n}{c_{n+1}}}}  \label{dcuatro}
\end{equation}
and use it to compute $c_{-1}/c_0$, starting with the approximate value
\begin{equation}
\frac{c_n}{c_{n+1}}\simeq 1+\frac{\sqrt{\omega a^2}}{n^{1/2}}\,, \qquad \mbox{for $n$ sufficiently large},  \label{dcinco}
\end{equation}
obtained from the characteristic equation \cite[Sec.1.6.3]{slav} of the recurrence relation.
The energy levels are obtained from the values of $Ea^2$ such that one gets
\begin{equation}
c_{-1}/c_0 = 0\,,    \label{dseis}
\end{equation}
which implies $c_{-1}=0$ and, in view of (\ref{dtres}), $c_{-2}=c_{-3}=\ldots=0$.

\section{Some results}

We have applied the above described algorithm to the determination of the four lowest states of a generalized harmonic oscillator with intensity $g$ varying from $g=0$ (harmonic oscillator) to $g=20$, for two typical values of $\omega a^2$, namely $\omega a^2=1/2$ and $\omega a^2=2$. The results are shown in Figures 1 and 2. The behaviour of the eigenenergies, as $g$ increases from zero, can be easily understood in view of the probability density of the harmonic oscillator eigenstates and the fact that the additional potential
\[
g\,\frac{x^2-a^2}{(x^2+a^2)^2}
\]
is negative for $|x|<a$ and positive for $|x|>a$ \cite[Fig. 1]{cari}. For low values of $g$, the effects on the energy of the positive and negative parts of that potential almost cancel to each other, except for the fundamental state whose density probility concentrates near the origin, where the additional potential is negative. This explains the gap between the energies of the fundamental and the first excited states encountered by Cari\~nena {\it et al.} \cite{cari} in the case of $\omega a^2 = 1/2$ and $g=2$. As $g$ increases further, the deeper and deeper potential well at the origin makes to decrease the energies of more and more excited states.

For possible numerical comparison with results obtained by other methods, we report ours, for some arbitrarily chosen values of $g$, in Tables 1 and 2. The energies of the four lowest states, given with ten decimal digits, have been obtained by using a double precision FORTRAN code. The presence, in Table 1, of exact values of the energy in the case of $g=2$ is not surprising: this is the exactly solvable case discussed by Cari\~nena {\it et al}. \cite{cari}. More intriguing are the exact values of $E_1a^2$ and $E_2a^2$ found in Table 1 for $g=12$ and those of $E_0a^2$ for $g=20$ and of $E_1a^2$ for $g=42$ in Table 2. As we are going to show in the next Section, these are also cases of quasi-polynomial wave functions. For those values of the parameters, one could speak of quasi-exactly solvable potentials.

\begin{figure}
\begin{center}
\epsfbox{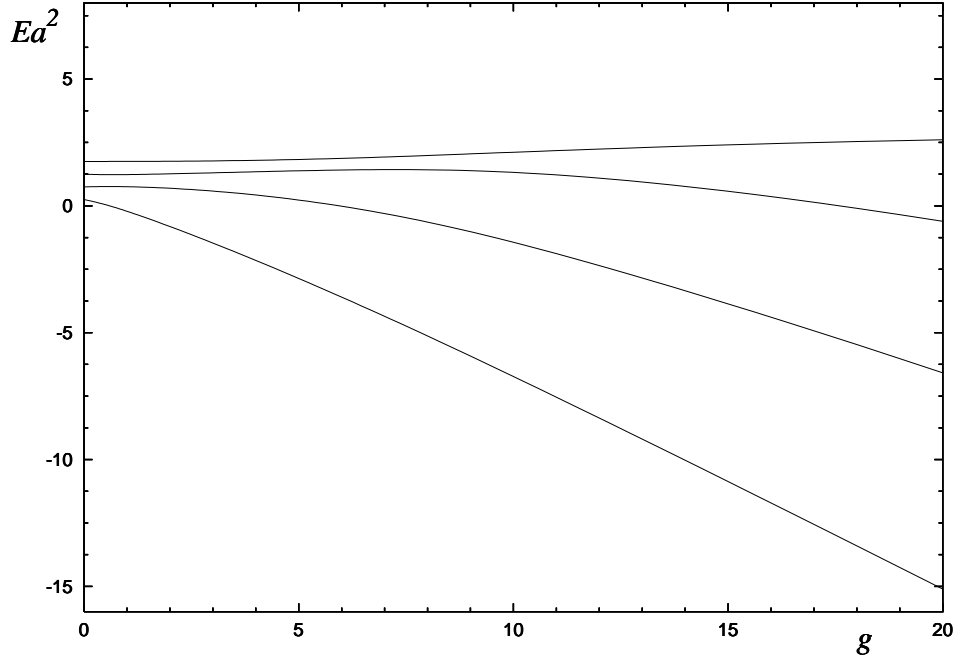}
\end{center}
\caption{Variation of the four lowest energy levels of the generalized isotonic oscillator with the intensity $g$. The parameters $\omega$ and $a$ of the oscillator are assumed to be such that $\omega a^2=1/2$.} \label{iso1}
\end{figure}

\begin{figure}
\begin{center}
\epsfbox{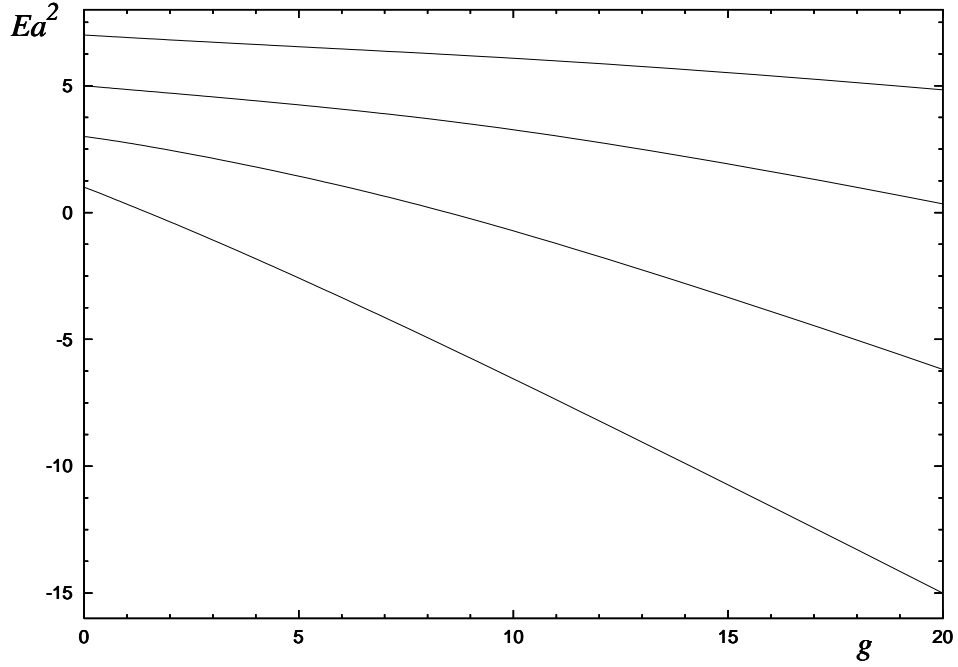}
\end{center}
\caption{Variation of the four lowest energy levels of the generalized isotonic oscillator with the intensity $g$, for $\omega a^2=2$.} \label{iso2}
\end{figure}

\begin{table}
\caption{Energies of the four lowest states  of the generalized isotonic oscillator of parameters $\omega$ and $a$ such that $\omega a^2=1/2$ and for some particular values of $g$.}
\begin{tabular}{lrrrr}
\hline
$g$ & $E_0a^2$ & $E_1a^2$ & $E_2a^2$ & $E_3a^2$  \\
\hline
0.00001 & 0.2499965567 & 0.7500003296 & 1.2499993408 & 1.7500000608 \\
0.1 & 0.2145449837 & 0.7531700367 & 1.2438631419  & 1.7505612654 \\
1 & $-$0.1849646064  & 0.7686368018  & 1.2241726194 & 1.7519905648 \\
2 & $-$0.75\phantom{00000000} & 0.75\phantom{00000000} & 1.25\phantom{00000000} & 1.75\phantom{00000000} \\
10 & $-$6.6939292571 & $-$1.3426308508 & 1.4335481682 & 2.1092577781 \\
12 & $-$8.3296211722& $-$2.25\phantom{00000000} & 1.25\phantom{00000000} & 2.2548656278 \\
20 & $-$15.1074137799 & $-$6.5820180239 & $-$0.6097201160 & 2.6043993900 \\
50 & $-$41.9395696487 & $-$27.0750778124 & $-$14.7363177369 & $-$4.9623193955 \\
\hline
\end{tabular}

\end{table}

\begin{table}
\caption{Energies of the four lowest states  of the generalized isotonic oscillator of parameters $\omega$ and $a$ such that $\omega a^2=2$ and for some chosen values of $g$.}
\begin{tabular}{lrrrr}
\hline
$g$ & $E_0a^2$ & $E_1a^2$ & $E_2a^2$ & $E_3a^2$  \\
\hline
0.00001 & 0.9999937095 & 2.9999977428 & 4.9999984646 & 6.9999989879 \\
0.1 & 0.9368657901 & 2.9772742737 & 4.9847133541  & 6.9898929491 \\
1 & 0.3495953307  & 2.7588911779  & 4.8519466428& 6.9003013951 \\
2 & $-$0.3372372644 & 2.4870257918 & 4.7099762556 &  6.8039923347 \\
5  & $-$2.5490351910 & 1.4941832183 & 4.2680431727 & 6.5346852493 \\
10 & $-$6.5291427792 & $-$0.6609393149 & 3.3184939783 & 6.1004000480 \\
12 & $-$8.1825461552& $-$1.6592922308 & 2.8380146272 & 5.9058815492 \\
20 & $-$15\phantom{.0000000000} & $-$6.1825461552 & 0.3407077692 & 4.8380146272 \\
42 & $-$34.5910045651 & $-$21\phantom{.0000000000} & $-$9.8551888685 & $-$1.1281494657 \\
50 & $-$41.8769597362 & $-$26.8630723075 & $-$14.3102873432 & $-$4.2061920738 \\
\hline
\end{tabular}
\end{table}

\section{Quasi-polynomial solutions}

The study of the generalized isotonic oscillator done by Cari\~nena {\it et al.} \cite{cari} revealed the existence of quasi-polynomial solutions for certain values of the parameters. Specifically, they found that, if the parameters are related in the form
\begin{equation}
g=2\omega a^2(2\omega a^2+1),   \label{dsiete}
\end{equation}
equation (\ref{uno}) admits a quasi-polynomial solution
\begin{equation}
\Psi_0=\frac{N_0}{(a^2+x^2)^{2\omega a^2}}\,\ \exp\left(-\frac{1}{2}\omega x^2\right), \qquad  E_0=\frac{1}{2}\omega-(2\omega a)^2,  \label{docho}
\end{equation}
$N_0$ being a normalization constant. It corresponds to the ground state of that generalized isotonic oscillator. More interestingly, they found that if, besides Eq. (\ref{dsiete}), one has
\begin{equation}
\omega a^2 = 1/2,  \label{dnueve}
\end{equation}
all eigenstates, of energies
\[
E_m= (m-3/2)\,\omega,
\]
are represented by quasi-polynomial wave functions
\begin{equation}
\Psi_m(x) = \frac{N_m}{\omega(x^2+a^2)}\exp\left(-\frac{\omega}{2} x^2\right)\mathcal{P}_m(\omega^{1/2}x), \qquad m=0, 3, 4, \ldots\,, \label{veinte}
\end{equation}
the polynomials $\mathcal{P}_m$ being linear combinations of three consecutive Hermite polynomials of the same parity.

The fact that simple analytic solutions of quantum problems, for particular values of the parameters, can serve as a check on numerical calculations was already pointed out by Demkov \cite{demk} in his study of the motion of a particle in the field of two Coulomb centers. He found that for certain set of values of the charges $Z_1$ and $Z_2$ of the centers and the distance $R$ between them, the wave function becomes quasipolynomial when written in ellipsoidal coordinates.

In order to analyze the possible  existence of quasi-polynomial solutions of Eq. (\ref{tres}), it is convenient to make explicit the behaviour of such solutions at the singular points. With this purpose we write
\begin{equation}
\Psi(z)=(z+1)^\rho\,z^\nu\,\exp\left(-\frac{\omega a^2}{2}\,z\right) \Phi(z), \label{vuno}
\end{equation}
where $\rho$ represents either $\rho_1$ or $\rho_2$, given by (\ref{cinco}) and such that
\[
\rho(\rho -1) = g\,,
\]
and, according to (\ref{seis}),  $\nu = \nu_1 =0$ in the case of even wavefunctions and $\nu =\nu_2=1/2$ for odd ones. The equation satisfied by $\Phi(z)$,
\begin{eqnarray}
\lefteqn{\frac{d^2\Phi}{dz^2} + \left(-\omega a^2+\frac{2\nu+1/2}{z}+\frac{2\rho}{z+1}\right)\,\frac{d\Phi}{dz}} \nonumber  \\
 & & \mbox{} + \frac{\omega a^2(\mu-\nu-\rho)z+\omega a^2(\mu-\nu)+\rho(2\nu+\rho/2)}{z(z+1)}\,\Phi=0\,,  \label{vdos}
\end{eqnarray}
with $\mu$ given in Eq. (\ref{ocho}), is again an example of CHE. In fact, the change of variable $z\longrightarrow -z$ permits to compare it with the non-symmetrical canonical form of the CHE \cite[Eq. (1.2.27)]{ronv}. Polynomial solutions of the CHE have been studied by Slavyanov \cite[Sec. 3.4]{ronv}. His analysis, however, is not directly applicable to our problem due to the different role played by the parameters in his equation and in ours.

It is possible to write a formal solution of (\ref{vdos}) as a power series
\begin{equation}
\Phi (z) = \sum_{n=0}^\infty a_{n}\,z^n\,, \label{vtres}
\end{equation}
with coefficients obeying the recurrence relation
\begin{equation}
\alpha_{n+2}\,a_{n+2} + \beta_{n+1}\,a_{n+1} + \gamma_n\,a_n=0,  \label{vcuatro}
\end{equation}
with
\[
 a_{-1}=0,   \qquad a_0\neq 0, \;\mbox{arbitrary},
\]
where we have abbreviated
\begin{eqnarray}
\alpha_m &=& m(m+2\nu-1/2)\,, \nonumber \\
\beta_m &=& m(m-\omega a^2 +2\nu+2\rho-1/2) +\omega a^2(\mu-\nu)+\rho(2\nu+\rho/2)\,, \nonumber \\
\gamma_m &=& -\omega a^2(m-(\mu-\nu-\rho))\,.  \nonumber
\end{eqnarray}
Obviously, the series in the right hand side of (\ref{vtres}) reduces to a polynomial
\begin{equation}
\Phi (z) = \mathcal{Q}_k = \sum_{n=0}^k a_{n,k}\,z^n\,, \label{vcinco}
\end{equation}
the coefficients $a_{n,k}$ being solution of (\ref{vcuatro}),
if the two conditions
\begin{equation}
\gamma_k=0   \qquad \mbox{and} \qquad a_{k+1,k}=0\label{vseis}
\end{equation}
are satisfied. Assuming that the parameters $\omega$ and $a^2$ are given, the first one of those conditions, gives the eigenenergy
\begin{equation}
E=(2k+2\rho+2\nu+1/2)\,\omega \label{vsiete}
\end{equation}
in terms of $\rho$, whereas the second one determines the values of $\rho$, and consequently of $g$, for which the quasi-polynomial (\ref{vuno}), with $\Phi(z)$ replaced by $\mathcal{Q}_k(z)$, is a solution. Notice that the second condition (\ref{vseis}) can be equivalently expressed as the cancelation of the determinant of a tridiagonal $(k+1)\times(k+1)$ matrix,
\begin{equation}
\det\left( \begin{array}{cccccc}
\beta_0 & \alpha_1 & & & & \\
\gamma_0 & \beta_1 & \alpha_2 & & & \\
 & \gamma_1 & \beta_2 & \alpha_3 & & \\
 & & \ddots & \ddots & \ddots & \\
 & & & \gamma_{k-2} & \beta_{k-1} & \alpha_k \\
 & & & & \gamma_{k-1} & \beta_k
 \end{array} \right) = 0\,,  \label{vocho}
 \end{equation}
with $\mu$ replaced by $k+\nu+\rho$ in the expressions of $\alpha_m$, $\beta_m$ and $\gamma_m$ given above. The left hand side of (\ref{vocho}) is a polynomial of degree $2(k+1)$ in $\rho$. One of its roots, for every value of $k$, is $\rho=0$: the pure harmonic oscillator possesses quasi-polynomial solutions of every degree. Obviously, conjugate pairs of complex roots may appear, but these are not interesting, in view of the restriction $g>0$. The real roots are easily obtained algebraically or numerically. To illustrate, we report the results obtained in the cases of $k=0, 1, 2$ and for the two values of $\omega a^2$ considered in Tables 1 and 2, respectively.

\subsection{Case k=0}

The first of conditions (\ref{vseis}), $\gamma_0=0$, gives $\mu=\nu+\rho$ and, in view of (\ref{ocho}),
\begin{equation}
E=(2\rho+2\nu+1/2)\,\omega. \label{vnueve}
\end{equation}
The second condition, $\beta_0=0$, gives
\begin{equation}
\rho\,(\omega a^2+2\nu+\rho/2)=0, \label{treinta}
\end{equation}
that, besides the trivial one $\rho=0$, has the solution
\begin{equation}
\rho=-(2\omega a^2 + 4\nu), \qquad \mbox{i. e.} \quad g=(2\omega a^2 + 4\nu)(2\omega a^2 + 4\nu+1)\,.  \label{tuno}
\end{equation}

\subsubsection{Even solutions ($\nu=0$).}

In the case of being
\[
\rho =-2\omega a^2, \quad \mbox{i. e.} \quad g=2\omega a^2(2\omega a^2+1),
\]
and for the energy
\[
E=(-4\omega a^2 +1/2)\,\omega,
\]
one has the quasi-polynomial solution
\[
\Psi(z) \propto(z+1)^{-2\omega a^2} \exp\left(-\frac{\omega a^2}{2}\,z\right)
\]
or, in terms of the original notation,
\[
\Psi(x) \propto(x^2+a^2)^{-2\omega a^2} \exp\left(-\frac{\omega}{2}\,x^2\right)\,.
\]
This is the solution mentioned in Eq. (2) of Ref. \cite{cari}. It can be recognized in our Table 1 (ground state for $g=2$) and in our Table 2 (ground state for $g=20$).

\subsubsection{Odd solutions ($\nu=1/2$).}

For
\[
\rho =-2\omega a^2-2, \quad \mbox{i. e.} \quad g=(2\omega a^2+2)(2\omega a^2+3),
\]
and energy
\[
E=(-4\omega a^2 -5/2)\,\omega,
\]
one finds the quasi-polynomial solution
\[
\Psi(z) \propto(z+1)^{-2\omega a^2-2}\, z^{1/2}\,\exp\left(-\frac{\omega a^2}{2}\,z\right)
\]
or, in terms of the variable $x$,
\[
\Psi(x) \propto(x^2+a^2)^{-2\omega a^2-2}\,x\,\exp\left(-\frac{\omega}{2}\,x^2\right)\,.
\]
Examples of the case under consideration appear in Tables 1 (first excited state for $g=12$) and 2 (first excited state for $g=42$).

\subsection{Case $k=1$}

\subsubsection{Even solutions ($\nu=0$).}

Now, from (\ref{vsiete}), we have
\[
E=\left(2\rho + 5/2\right)\,\omega\,.
\]
and, from (\ref{vocho}),
\[
\frac{1}{4}\,\rho \left(\rho^3+4(\omega a^2+1)\rho^2+(4\omega^2a^4+10\omega a^2+1)\rho+2\omega a^2(2\omega a^2+5)\right) = 0\,,
\]
For $\omega a^2=1/2$ there is a nontrivial real solution
\[
\rho = -A^{1/3}/3-2-5\,A^{-1/3}, \qquad \mbox{with} \quad A=3\left(36-\sqrt{921}\right),
\]
corresponding to an intensity of the additional potential
\[
g = A^{2/3}/9+5A^{1/3}/3+28/3+25A^{-1/3}+25A^{-2/3}\,.
\]
For $\omega a^2=2$ there is also one nontrivial real solution
\[
\rho  = -B^{1/3}/3 -4-11\,B^{-1/3}, \qquad \mbox{with} \quad B=3\left(72-\sqrt{1191}\right),
\]
to which it corresponds
\[
g = B^{2/3}/9+3B^{1/3}+82/3+99B^{-1/3}+121B^{-2/3}\,.
\]

\subsubsection{Odd solutions ($\nu=1/2$).}

For energy we have the value
\[
E=\left(2\rho + 7/2\right)\,\omega\,.
\]
and for $\rho$ the equation
\[
\frac{1}{4}\,\rho \left(\rho^3+4(\omega a^2+2)\rho^2+(4\omega^2a^4+18\omega a^2+15)\rho+2(2\omega^2 a^4+9\omega a^2+3)\right) = 0\,,
\]
For $\omega a^2=1/2$ one gets three nontrivial real solutions, namely,
\[
\rho =  -\, 1\,,    \qquad
\rho =  -\, \frac{1}{2}\left(9-\sqrt{17}\right)\,, \qquad
\rho =  -\, \frac{1}{2}\left(9+\sqrt{17}\right)\,,
\]
with corresponding intensities
\[
g =  2\,,    \qquad
g =  29-5\,\sqrt{17}\,,    \qquad
g =  29+5\,\sqrt{17}\,.
\]

For $\omega a^2=2$ one obtains also three nontrivial real solutions
\begin{eqnarray}
\rho & =& -\, (1/3)\left(16-55^{1/3}\cdot 2\,\Re\left(\left(-1+3i\sqrt{6}\right)^{1/3}\right)\right)\,,    \nonumber  \\
\rho & =& -\, (1/3)\left(16+55^{1/3}\,\Re\left(\left(1+i\sqrt{3}\right)\left(-1+3i\sqrt{6}\right)^{1/3}\right)\right)\,,    \nonumber  \\
\rho & =& -\, (1/3)\left(16+55^{1/3}\,\Re\left(\left(1-i\sqrt{3}\right)\left(-1+3i\sqrt{6}\right)^{1/3}\right)\right)\,.    \nonumber
\end{eqnarray}
The corresponding values of $g$ are obtained by using
\begin{equation}
g =\rho(\rho-1)\,.   \label{extra}
\end{equation}

\subsection{Case $k=2$}

\subsubsection{Even solutions ($\nu=0$).}

The energy is now given by
\[
E=\left(2\rho + 9/2\right)\,\omega\,,
\]
where $\rho$ is a solution of
\begin{eqnarray*}
\lefteqn{\frac{1}{8}\,\rho \Big(\rho^5  +6(\omega a^2+2)\rho^4+3(4\omega^2a^4+18\omega a^2+13)\rho^3}  \\
& & \mbox{} + 4(2\omega^3 a^6+18\omega^2 a^4+39\omega a^2+8)\rho^2 + 2(12\omega^3 a^6+82\omega^2 a^4+108\omega a^2+3)\rho   \\
& & \mbox{} +4(4\omega^3 a^6+28\omega^2 a^4+27\omega a^2)\Big) = 0\,, 
\end{eqnarray*}
For $\omega a^2=1/2$, this equation has, besides the trivial solution, two complex and the real ones
\begin{eqnarray*}
\rho & = & -\,1\,,   \\
\rho & = & -\,\frac{1}{2}\left(7+C^{1/2}-\left(37-C+64\,C^{-1/2}\right)^{1/2}\right)\,,   \\
\rho & = & -\,\frac{1}{2}\left(7+C^{1/2}+\left(37-C+64\,C^{-1/2}\right)^{1/2}\right)\,,  
\end{eqnarray*}
where we have abbreviated
\[
C=\frac{1}{3}\left(37+5^{1/3}\left(\left(7967-6\,\sqrt{883749}\right)^{1/3}+
\left(7967+6\,\sqrt{883749}\right)^{1/3}\right)\right)\,.
\]
The corresponding intensities of the additional potential are given by Eq. (\ref{extra}).

For $\omega a^2=2$ there are also three non trivial real solutions,
\[
\rho\simeq -1.0813491830140753\,,\quad \rho\simeq -7.22694217991115\,, \quad \rho\simeq -11.525122115065377\,,
\]
which correspond to
\[
 g\simeq 2.2506652386192836\,, \quad g\simeq 59.45563545169008\,, \quad g\simeq 144.3535618822344\,.
\]

\subsubsection{Odd solutions ($\nu=1/2$).}

We have for the energy 
\[
E=\left(2\rho + 11/2\right)\,\omega\,,
\]
and for $\rho$
\begin{eqnarray*}
\lefteqn{\frac{1}{8}\,\rho \Big(\rho^5 +6(\omega a^2+3)\rho^4+3(4\omega^2a^4+26\omega a^2+35)\rho^3}  \\ 
& & \mbox{} + 4(2\omega^3 a^6+24\omega^2 a^4+81\omega a^2+59)\rho^2+2(12\omega^3 a^6+118\omega^2 a^4+276\omega a^2+105)\rho  \\
& &\mbox{} + 4(4\omega^3 a^6+40\omega^2 a^4+75\omega a^2+15)\Big) = 0\,. 
\end{eqnarray*}
For $\omega a^2=1/2$ we obtain three nontrivial real solutions,
\begin{eqnarray*}
\nu & = & -\, 1\,,     \\
\nu & = & \,5- \frac{1}{2}\left(D^{1/2}-\left(46-D+52\,D^{-1/2}\right)^{1/2}\right)\,,      \\
\nu & = & -\,5- \frac{1}{2}\left(D^{1/2}+\left(46-D+52\,D^{-1/2}\right)^{1/2}\right)\,,   
\end{eqnarray*}
where
\[
D=\frac{1}{3}\left(46+\left(56575-12\,\sqrt{3456147}\right)^{1/3}+
\left(56575+12\,\sqrt{3456147}\right)^{1/3}\right)\,.
\]
The corresponding values of $g$ follow from Eq. (\ref{extra}).

For $\omega a^2=2$ there are also three nontrivial real solutions,
\[
 \rho\simeq -1.0527990898606965\,, \quad \rho\simeq -9.18056490570385\,, \quad \rho\simeq -13.30773725973479\,,
\]
corresponding to intensities
\[
 g\simeq 2.1611850134722075\,, \quad g\simeq 93.46333689354499\,, \quad g\simeq 190.40360823493683\,.
\]

None of the above considered cases explains the value $E_2a^2=5/4$ found for $\omega a^2=1/2$ (Table 1) and $g=12$. It is not difficult to see that it corresponds to $k=4$, $\nu=0$, and $\rho =-3$. The wave function is in this case
\[
\Psi (z)\propto (z+1)^{-3}\exp(-z/4)\left(1-10\,z-4\,z^2-(2/3)z^3-(1/21)z^4\right)\,,
\]
that is, in terms of $x$,
\[
\Psi (x)\propto (x^2+a^2)^{-3}\exp(-x^2/4a^2)\left(1-\frac{10\,x^2}{a^2}-\frac{4\,x^4}{a^4}-\frac{2\,x^6}{3\,a^6}-\frac{x^8}{21\,a^8}\right)\,.
\]

\section{The exactly solvable case}

In the preceding Section we have found, whenever $\omega a^2=1/2$, a quasi-polynomial solution with $\rho=-1$ for almost every considered value of $k$ and in both cases of even or odd wave functions. There are only two exceptions, namely the case of $k=0$, $\nu=1/2$ and that of $k=1$, $\nu=0$. Although we have not discussed the cases of $k=3, 4, \ldots $, one can easily check that the same value $\rho=-1$ appears for $\nu=0$ and for $\nu=1/2$ in all cases. This was to be expected, since for $\omega a^2=1/2$ and $\rho=-1$, that is, $g=2$, the generalized isotonic potential is exactly solvable \cite{cari}. In what follows, we show that the polynomials entering the solutions found by Cari\~nena {\it et al.} are in fact confluent Heun polynomials.

We assume from now on that $\omega a^2=1/2$, $\rho=-1$, and $\mu=k+\nu-1$, $k$ being a positive integer. According to Eqs. (\ref{vuno}), (\ref{vdos}) and (\ref{vcinco}), the quasi-polynomial solutions are of the form
\begin{equation}
\Psi(z)=(z+1)^{-1}\,z^\nu\,\exp\left(-z/4\right) \mathcal{Q}_k^{(\nu)}(z)\,, \qquad \nu = 0,\, 1/2\,, \label{tdos}
\end{equation}
the polynomial (of degree $k$) $\mathcal{Q}_k^{(\nu)}$ obeying the differential equation
\begin{equation}
\frac{d^2\mathcal{Q}_k^{(\nu)}}{dz^2} + \left(-\frac{1}{2}+\frac{2\nu+1/2}{z}-\frac{2}{z+1}\right)\,\frac{d\mathcal{Q}_k^{(\nu)}}{dz}
+ \frac{kz/2+k/2-2\nu}{z(z+1)}\,\mathcal{Q}_k^{(\nu)}=0\,.  \label{ttres}
\end{equation}
This is but a particular case of Eq. (\ref{vdos}) that, as we have already mentioned, is a CHE. Confluent Heun polynomials can be written as linear combinations of hypergeometric or confluent hypergeometric polynomials \cite[Sec. 2.3]{ronv}. This second possibility is more convenient for a comparison with the results of Cari\~nena {\it et al.} \cite{cari}. With this purpose, we introduce a new variable
\begin{equation}
y=\frac{z}{2}=\frac{x^2}{2a^2}\,,   \label{tcuatro}
\end{equation}
in terms of which the differential equation reads
\begin{eqnarray}
 \lefteqn{\hspace{-3cm}(y\! +\! 1/2)y\,\frac{d^2\mathcal{Q}_k^{(\nu)}}{dy^2} + \left((y\! +\! 1/2)(2\nu\! +\! 1/2\! -\! y)-2y\right)\frac{d\mathcal{Q}_k^{(\nu)}}{dy}} \nonumber  \\
& & \hspace{1cm}\mbox{} + \left((y\! +\! 1/2)k-2\nu\right)\mathcal{Q}_k^{(\nu)}=0\,,  \label{tcinco}
\end{eqnarray}
that can be written in the form
\begin{equation}
\left((y+1/2)\mathcal{D}_0 + \mathcal{D}_1\right)\,\mathcal{Q}_k^{(\nu)} = 0  \label{tseis}
\end{equation}
with the differential operators
\begin{eqnarray}
\mathcal{D}_0 & \equiv & y\,\frac{d^2}{dy^2}+(2\nu+1/2-y)\frac{d}{dy} + k\,,  \label{tsiete}  \\
\mathcal{D}_1 & \equiv & -2y\,\frac{d}{dy} - 2\nu\,.  \label{tocho}
\end{eqnarray}
Now we try in (\ref{tseis}) the sum of confluent hypergeometric polynomials
\begin{equation}
\mathcal{Q}_k^{(\nu)}=\sum_{n=0}^k\mathcal{A}_{n,k}^{(\nu)}\,M(-(k-n), 2\nu+1/2, y)\,,  \label{tnueve}
\end{equation}
with coefficients $\mathcal{A}_{n,k}^{(\nu)}$ to be determined. According to Eqs. 13.1.1  and  13.4.10, respectively, of Ref \cite{abra}, one has
\begin{eqnarray}
\mathcal{D}_0\, M(-(k-n), 2\nu+1/2, y) & =& n\,M(-(k-n), 2\nu+1/2, y)\,,  \label{cuarenta}  \\
\mathcal{D}_1\, M(-(k-n), 2\nu+1/2, y) & =&-2(k-n+\nu)\,M(-(k-n), 2\nu+1/2, y)  \nonumber  \\
 & & \hspace{-1cm}\mbox{}+2(k-n)\,M(-(k-n)+1, 2\nu+1/2, y)\,,  \label{cuno}
\end{eqnarray}
and Eq. (\ref{tseis}) turns into
\begin{eqnarray}
\lefteqn{\sum_{n=0}^k\mathcal{A}_{n,k}^{(\nu)}\, \Big(\left( n(y+5/2)-2k-2\nu\right)M(-(k-n), 2\nu+1/2, y)}  \nonumber  \\
& & \hspace{1cm} \ +2(k-n)M(-(k-n)+1, 2\nu+1/2, y)\Big)=0\,. \label{cdos}
\end{eqnarray}
Cancelation of the coefficients of the successively decreasing powers of $y$ in the left hand side of the last equation gives, for $k\geq 2$,
\begin{eqnarray*}
\mathcal{A}_{0,k}^{(\nu)} & &\quad\mbox{arbitrary}\,,   \\
\mathcal{A}_{1,k}^{(\nu)} & = & -2\,\frac{k+\nu}{k+2\nu-1/2}\,\mathcal{A}_{0,k}^{(\nu)}\,,  \\
\mathcal{A}_{2,k}^{(\nu)} & = & \frac{2k^2-3k+4k\nu-2\nu}{2(k+2\nu-1/2)(k+2\nu-3/2)}\,\mathcal{A}_{0,k}^{(\nu)}\,,  \\
\mathcal{A}_{3,k}^{(\nu)} & =  & \mathcal{A}_{4,k}^{(\nu)} = \ldots = 0\,, 
\end{eqnarray*}
in both cases of $\nu=0$ or $\nu=1/2$. Substitution of these expressions in (\ref{tnueve}) gives
\begin{equation}
\mathcal{Q}_k^{(0)} = \mathcal{A}_{0,k}^{(0)}\left(M(-k,1/2,y)-\frac{2k}{k-\frac{1}{2}}M(-k\! +\! 1,1/2,y)+\frac{k}{k-\frac{1}{2}}M(-k\! +\! 2,1/2,y)\right), \label{ctres}
\end{equation}
\begin{equation}
\mathcal{Q}_k^{(1/2)} = \mathcal{A}_{0,k}^{(1/2)}\left(M(-k,3/2,y)-2\,M(-k\! +\! 1,3/2,y)+\frac{k-1}{k-\frac{1}{2}}M(-k\! +\! 2,3/2,y)\right). \label{ccuatro}
\end{equation}
Now we can use the relations  between confluent hypergeometric and Hermite polynomials \cite[Eqs. 22.5.56 and 22.5.57]{abra} \cite[Sec. 10.13, Eqs. (17) and (18)]{erde}
\begin{equation}
M(-m,1/2,y) = (-1)^m\frac{m!}{(2m)!}\,H_{2m}\left(\sqrt{y}\right)\,,  \label{ccinco}
\end{equation}
\begin{equation}
\sqrt{y}\,M(-m,3/2,y) = (-1)^m\frac{m!}{2(2m+1)!}\,H_{2m+1}\left(\sqrt{y}\right)\,,  \label{cseis}
\end{equation}
and choose
\begin{equation}
\mathcal{A}_{0,k}^{(0)}=(-1)^k\,\frac{(2k)!}{k!} \qquad \mbox{and} \qquad  \mathcal{A}_{0,k}^{(1/2)}=(-1)^k\,\frac{2^{1/2}(2k+1)!}{k!}\,  \label{csiete}
\end{equation}
to get
\begin{equation}
\mathcal{Q}_k^{(0)} = H_{2k}(\sqrt{y})+8kH_{2k-2}(\sqrt{y})+8k(2k-3)H_{2k-4}(\sqrt{y})\,, \label{cocho}
\end{equation}
\begin{equation}
 z^{1/2}\,\mathcal{Q}_k^{(1/2)} = H_{2k+1}(\sqrt{y})+4(2k+1)H_{2k-1}(\sqrt{y})+4(2k+1)(2k-2)H_{2k-3}(\sqrt{y})\,, \label{cnueve}
\end{equation}
which are, respectively, the polynomials $\mathcal{P}_{2k}$ and $\mathcal{P}_{2k+1}$ of Ref. \cite{cari}, the variable being
\begin{equation}
\sqrt{y} = \frac{x}{\sqrt{2}\,a} = \sqrt{\omega}\, x\,.  \label{cincuenta}
\end{equation}

In the above discussion of Eq. (\ref{cdos}) we have left aside the cases of $k=0$ and $k=1$. For $k=0$ and $\nu=0$, Eq. (\ref{cdos}) is satisfied irrespective of the value of $\mathcal{A}_{0,0}^{(0)}$, whereas for $\nu =1/2$ its fulfilment requires $\mathcal{A}_{0,0}^{(1/2)}=0$, which implies that the resulting $\mathcal{Q}_{0}^{(1/2)}$ is identically equal to zero. For $k=1$, instead, Eq. (\ref{cdos}) is satisfied, for $\nu =0$, only if $\mathcal{A}_{0,1}^{(0)} = \mathcal{A}_{1,1}^{(0)} = 0$ and consequently $\mathcal{Q}_{1}^{(0)}$ identically equal to zero, whereas for $\nu=1/2$ it only requires $\mathcal{A}_{1,1}^{(1/2)} = -2\,\mathcal{A}_{0,1}^{(1/2)}$. All this is in accordance with the two exceptional cases of $k=0$, $\nu=1/2$, and $k=1$, $\nu=0$, encountered in Ref. \cite{cari} and mentioned at the beginning of this Section.

\section{Final comments}

Most of the quantum mechanical problems which admit a simple solution are of the hypergeometric class: the corresponding Schr\"odinger equation turns, by an adequate transformation, into a hypergeometric or a confluent hypergeometric one. In this paper we have shown that the generalized quantum isotonic oscillator belongs to the Heun class. Other examples of physical problems of this class can be found in Chapter 4 of Ref. \cite{slav}.

Up to now, the most appealing  feature of the generalized quantum isotonic potential was its exact solvability for certain values of its parameters \cite{cari}. The fact, shown in this paper, that it is quasi-exactly solvable when those parameters take particular values spread over a wide range makes it to be especially suited to serve as a workbench to test the accuracy of approximate (perturbative, variational, etc.) methods of solution of the Schr\"odinger equation.

Quasi-exact solvability of quantum Hamiltonians is closely related to their Lie-algebraic properties \cite{gonz}. A discussion of this topic with reference to the generalized isotonic oscillator would be necessary, but it lies out of the scope of this paper.

The sequence $\{\mathcal{P}_n\}$ arising in the exactly solvable case \cite{cari} does not include a linear ($n=1$) nor a quadratic ($n=2$) polynomials. Therefore, it cannot be used as a basis for an expansion. Similar sequences of polynomial eigenfunctions of a Sturm-Liouville problem have been found by G\'omez-Ullate, Kamran and Milson \cite{gom1} and by Quesne \cite{ques}. In the case of the first authors, the sequences does not include the constant ($n=0$) polynomial. Quesne has found sequences without the constant polynomial and also sequences without the constant and the linear polynomials. Nevertheless, the first authors have proved that such sequences, which they denominate {\it exceptional polynomial systems}, are a basis in their corresponding $L^2$ Hilbert spaces. Besides, G\'omez-Ullate, Kamran and Milson \cite{gom2} have given an extension of the Bochner's theorem applicable to those sequences of orthogonal polynomials, solution of a Sturm-Liouville problem, that start with a polynomial of degree one. It would be interesting to explore the possibility of an analogous extension of the theorem for sequences of polynomials like that discussed in Ref. \cite{cari},
where the absent polynomials are not the lowest order ones.

\section*{Acknowledgments}

 The author is greatly indebted to Professors J. F. Cari\~nena and M. F. Ra\~nada for providing him with their paper and for plenty of fruitful comments. The suggestions of two anonymous referees have greatly contributed to improve the presentation of this article. Financial aid of Comisi\'on Interministerial de Ciencia y Tecnolog\'{\i}a and of Diputaci\'on General de Arag\'on is acknowledged.

\end{document}